\documentclass[12pt,english,aps,manuscript]{article}
\usepackage[T1]{fontenc}
\usepackage[latin1]{inputenc}
\usepackage{geometry}
\usepackage[active]{srcltx}
\usepackage{amsmath}
\usepackage{amssymb}
\usepackage[numbers]{natbib}

\makeatletter
\usepackage{geometry}

\makeatletter
\usepackage{color}

\makeatletter
\newcommand{\bee}{\begin{equation}}
\newcommand{\ee}{\end{equation}}
\newcommand{\beea}{\begin{eqnarray}}
\newcommand{\eea}{\end{eqnarray}}

\makeatother

\makeatother

\makeatother

\usepackage{babel}
\begin{document}
\begin{center}
\textbf{\Large{}The Wave Function of the Universe and CMB Fluctuations }
\par\end{center}{\Large \par}

\begin{center}
\vspace{0.3cm}

\par\end{center}

\begin{center}
{\large{}S. P. de Alwis$^{\dagger}$ }
\par\end{center}{\large \par}

\begin{center}
Physics Department, University of Colorado, \\
 Boulder, CO 80309 USA 
\par\end{center}

\begin{center}
\vspace{0.3cm}
 
\par\end{center}

\begin{center}
\textbf{Abstract} 
\par\end{center}

\begin{center}
\vspace{0.3cm}

\par\end{center}

The Hartle-Hawking and Tunneling (Vilenkin) wave functions are treated
in the Hamiltonian formalism. We find that the leading (i.e. quadratic)
terms in the fluctuations around a maximally symmetric background,
are indeed Gaussian (rather than inverse Gaussian), for both types
of wave function, when properly interpreted. However the suppression
of non-Gaussianities and hence the recovery of the Bunch-Davies state
is not transparent.

\smallskip{}
\vspace{0.3cm}

\vfill{}

$^{\dagger}$ dealwiss@colorado.edu 

\eject

\section{Introduction}

Inflationary cosmology is not past eternal. Under the assumption that
the average Hubble parameter $<H>>0$, i.e. that on average the universe
has been expanding in the past, both null and time-like geodesics
cannot be arbitrarily extended to the past \citep{Borde:2001nh}.
In fact unlike the cosmological singularity theorems which relied
on the use of the weak energy condition (violated by inflationary
cosmology), this argument does not need such a condition.

On the other hand the standard calculation of the scalar and tensor
fluctuation spectrum around the inflationary background assumes that
in the far past (conformal time $\eta\rightarrow-\infty(1-i\epsilon)$),
the inflationary background remains valid, and correlation functions
in any state of the system (defined in the interaction picture) at
some conformal time $\tau$, is related to those in the ``Bunch-Davies''
(BD) vacuum state $|0>$ by the ``in-in'' formula \citep{Maldacena:2002vr,Weinberg:2005vy},
\begin{eqnarray}
\frac{<\Omega(\tau)|\hat{W}(\tau)|\Omega(\tau)>}{<\Omega(\tau)|\Omega(\tau)>} & = & <0|U^{\dagger}(\tau,-\infty)W(\tau)U(\tau,-\infty)|0>,\label{eq:inin}\\
U(\tau,-\infty) & = & T\exp\left(-\frac{i}{\hbar}\int_{-\infty(1-i\epsilon)}^{\tau}\hat{H}_{1I}(\tau')d\tau'\right).
\end{eqnarray}
In the above the Hamiltonian for fluctuations around the (time-dependent)
inflationary background is taken to be of the form $H_{0}(t)+H_{1}(t)$,
where the first term governs quadratic fluctuations and the second
cubic and higher order fluctuations. Also $\hat{H}_{1I}$ is $H_{1}$
in the interaction picture. 

The important point to note here is that both $H_{0}$ and $H_{1}$
are time dependent, with this dependence given by the slow roll inflationary
background metric. The projection onto the matrix element in the BD
vacuum is a consequence of taking the limit of infinite negative conformal
time. This however seems to be in apparent conflict with the theorem
of \citep{Borde:2001nh} that inflation is not past eternal. 

This motivates the search for some explanation as to the emergence
of an inflationary background starting from ``nothing''. This has
been discussed since the early eighties and there are two main proposals.
One is the ``no boundary'' wave function of Hartle and Hawking \citep{Hartle:1983ai}
(HH). The other is the tunneling (T) wave function of Vilenkin \citep{Vilenkin:1982de}\footnote{For a recent discussion and references to the early literature see
\citep{Vilenkin:2018dch}. This paper also addresses similar issues
to those in this paper but only for the T wave function and from a
different perspective. See also \citep{Kamenshchik:2018pwf} and references
therein for a related discussion, as well as \citep{Bojowald:2018gdt}
for a loop quantum gravity perspective.}. While both wave functions could in principle be valid solutions
to the Wheeler DeWitt equation without truncation - in practice the
explicit forms of the wave functions have only been obtained in the
so-called mini-superspace model where only the temporal dependence
of the fields is kept.

Recently this basis for justifying the BD state in the inflationary
background has been questioned \citep{Feldbrugge:2017kzv,Feldbrugge:2017fcc,Feldbrugge:2017mbc,Feldbrugge:2018gin}.
Using the Picard-Lefshetz theory of saddle point approximations to
the integral over the lapse $N$ of the ADM formulation of general
relativity, it was claimed in \citep{Feldbrugge:2017kzv} that this
method justifies the tunneling wave function but not the HH wave function.
In the subsequent paper \citep{Feldbrugge:2017fcc}, the authors then
argued that quadratic scalar/tensor fluctuations around both T and
HH wave functions were unsuppressed, having an inverse Gaussian form.
This seemed to cast doubt on the quantum cosmology justification for
a smooth BD beginning to inflationary cosmology. 

In a rebuttal of these claims one of the original proponents of the
HH wave function (Hartle) and collaborators \citep{DiazDorronsoro:2017hti},
argued that there is a choice of contour that justifies the HH wave
function. Furthermore, whereas the T wave function (following from
the contour of \citep{Feldbrugge:2017kzv}) was indeed unstable due
to unsuppressed fluctuations (as argued a while ago also in the references
quoted in \citep{DiazDorronsoro:2017hti}), the HH wave function is
not. Subsequently in \citep{Feldbrugge:2017mbc} it was argued that
the contour chosen in \citep{DiazDorronsoro:2017hti} actually should
pick up subdominant saddle points which restores the unsuppressed
fluctuations in the HH wave function as well.

A new round of claims and counter claims were made this year (2018).
Reference \citep{DiazDorronsoro:2018wro} considered a generalization
of minisuperspace, replacing the round $S_{3}$ by axial Bianchi IX
geometry. They argued, using a circular contour for the integration
over the lapse $N$, that the HH wave function is well defined and
hence that the original HH wave function was stable under deformations.
This choice of contour was criticized in \citep{Feldbrugge:2018gin}
on the grounds that it was not physically well motivated and leads
to ``mathematical and physical inconsistencies''. Finally in a very
recent paper Vilenkin and Yamada \citep{Vilenkin:2018dch} have argued
that provided certain boundary/initial conditions on the scalar fluctuations
are satisfied, the scalar field fluctuations around the tunneling
wave function (T) are well behaved.

Our point of view is that quantum gravity should be defined in terms
of the Wheeler De Wit (WdW) equation - which is a constraint equation
whose solutions are possible wave functions of the universe. The functional
integral definition of the wave function is simply a method for computing
it\footnote{This appears to be the position of one of the founding fathers of
quantum cosmology and his collaborators as well - see for instance
the recent paper \citep{Halliwell:2018ejl} which appeared after the
first version of this paper was posted.}. But there are equally valid approaches to solving the WdW equation
such as the WKB method. In this paper we will use the latter method
to discuss the possible solutions to the WdW equation in the semi-classical
approximation and rederive solutions that had been obtained before
using the functional integral method. Including the fluctuations around
the cosmological deSitter background we find four independent solutions.
Different choices of contour on the two sides of this dispute just
correspond to different choices of integration constants. We do not
believe that there is any ``fundamental principle'' that dictates
one or other choice as seems to be the position of Feldebrugge et
al.. In fact as they point out, their contour leads to unsuppressed
- i.e. inverse Gaussian - fluctuations - in contradiction to what
is observed. On the other hand as shown in \citep{DiazDorronsoro:2017hti,DiazDorronsoro:2018wro}
there exists an alternate choice of contour which leads to the HH
wave function with suppressed fluctuations. 

This paper is aimed at reconciling these different points of view.
Our main argument is that the WdW equation implies that there is no
notion of time (and therefore of time ordering). It is essentially
like the time independent Schroedinger equation whose solutions are
stationary states. This of course is the famous ``problem of time''
in quantum gravity. One can compare the probabilites for different
configurations of the fields of the system and its geometry. But there
is no notion of which configuration is prior to which.

Thus in the present context there are 4 different particular solutions
of the WdW equations in the semi-classical approximation, both inside
and outside the effective potential barrier, which can then be matched
as in the standard WKB procedure. By appropriate choices of integration
constants, these solutions can be organized either into tunneling
type wave functions or into HH type (real) wave functions, both with
suppressed Gaussian fluctuations. On the other hand one also has (for
both cases) solutions with unsuppressed (inverse Gaussian) fluctuations.
These are the solutions that were obtained by Feldebrugge et al. from
their integration contour for the lapse. Clearly these are unphysical
(certainly inconsistent with the observed smooth homogenous isotropic
universe) and hence should be rejected. 

We will first discuss two simple problems which illustrate one of
the main points which we wish to make, namely that there is no advantage
to using the functional integral and the Picard - Lefshetz method
to solve the WdW equation wherever it is possible to solve the equation
directly in the semi-classical approximation, which is the case in
all these examples. In fact in the simplest example, that of one-particle
non-relativistec quantum mechanics, derived from a time-reparametrization
invariant action leading to a WdW equation, it is clear that while
one can indeed define the semi-classical solution in terms of an integral
over the lapse - the dependence on the latter drops out of the classical
action when one uses the constraint equation. Next we discuss the
Schwinger process where essentially the same is true. In effect we
argue that one needs some physical input to decide what particular
solution to pick, and that one cannot do this on mathematical grounds
or on the basis of an a priori notion of causality. We then review
the mini-superspace solution for universe creation from ``nothing''
- both the HH and the T cases. Finally we discuss fluctuations around
mini-superspace. We will find that, properly interpreted, both HH
and T cases lead to suppressed Gaussian fluctuations for the wave
function in the classical regime, contrary to the claims in \citep{Feldbrugge:2017fcc,Feldbrugge:2017mbc,Feldbrugge:2018gin}.
On the other hand this wave function (in both HH or T cases) necessarily
has non-Gaussian (i.e. cubic and higher powers of fluctuations) terms.
In other words the emergence of the Bunch-Davies vacuum wave function,
which is just Gaussian, does not appear to have any explanation from
these considerations.

\section{Quantum mechanics Examples}

In the following we will discuss two examples which show that both
the Hamiltonian argument and the path integral one agree and that
the latter does not resolve the ambiguity involved in the choice of
wave function.

\subsection{Particle in a potential}

We consider a 1D non-relativistic particle action with time reparametrization
invariance.
\begin{equation}
S=\int_{0}^{1}dt[N(t)^{-1}\frac{1}{2}\dot{x}^{2}-N(t)(V-E)].\label{eq:NRS}
\end{equation}
\begin{eqnarray}
p_{x} & = & \frac{\dot{x}}{N},\,\,p_{N}=0,\label{eq:NRp}\\
{\cal H} & = & p_{x}\dot{x}-L=N(\frac{1}{2}p^{2}+V-E).\label{eq:NRH}
\end{eqnarray}
We have the secondary constraint
\[
\dot{p}_{N}=\{p_{N,}{\cal H}\}=(\frac{1}{2}p_{x}^{2}+V-E)=0,
\]
i.e. the Hamiltonian is weakly zero
\[
{\cal H}\approx0.
\]
This is of course completely equivalent to the usual energy conservation
equation. Also we have 
\begin{equation}
\dot{N}=\{N,{\cal H}\}=0.\label{eq:Ndot}
\end{equation}
 So we can choose the gauge $N={\rm constant}$. The (time independent)
Schroedinger equation is then the same as the Wheeler DeWitt equation,
i.e. with $p_{x}\rightarrow\hat{p}_{x}=-i\hbar\frac{d}{dx}$ 
\begin{equation}
\hat{H}\Psi(x)=\left(-\frac{\hbar^{2}}{2}\frac{d^{2}}{dx^{2}}+V(x)-E\right)\Psi(x)=0.\label{eq:WdWS}
\end{equation}
Solving this by putting $\Psi[q,\phi]=e^{\frac{i}{\hbar}S[q,\phi]}$
with $S=S_{0}+\hbar S_{1}$ gives 
\[
\frac{1}{2}\left(\frac{dS_{0}}{dx}\right)^{2}+V-E=0,\,\,\frac{dS_{1}}{dx}=\frac{i}{2}\frac{d}{dx}\ln\frac{dS_{0}}{dx}.
\]
This gives in the semi-classical approximation, the usual WKB result
\begin{equation}
\Psi(x)=\frac{c}{|2(E-V(x))|^{1/4}}e^{\pm\frac{i}{\hbar}\int^{x}\sqrt{2(E-V(x'))}dx'},\label{eq:PsiWKB}
\end{equation}
and clearly both solutions are allowed. In the regime where $V(x)>E$
we have both an exponentially rising and a falling solution with absolutely
no mathematical reason for discarding one or the other. Similarly
(assuming that the potential is greater than zero only over a finite
interval (so that what we have is a barrier rather than an infinite
wall) one has outgoing and incoming solutions, again with no reason
to discard either.

Now consider the path integral for the Feynman Kernel for this problem.
The Feynman kernel to go from $x_{0}$ at time $t_{0}=0$ to $x_{1}$
at $t_{1}=1$ is (note $K$ is a function of the first four variables
and a functional of $N$)
\begin{eqnarray}
K(x_{1},t_{1};x_{0},t_{0}:N] & = & <x_{1}|Te^{-\frac{i}{\hbar}\int_{t_{0}}^{t_{1}}{\cal \hat{H}}dt}|x_{0}>\nonumber \\
 & = & \int\prod_{t_{0}}^{t_{1}}dp(t_{i})\int\prod_{x(t_{0})=x_{0}}^{x(t_{1})=x_{1}}dx(t_{i})e^{\frac{i}{\hbar}\int_{t_{0}}^{t_{1}}dt[p(t)\dot{x}(t)-N(t)(\frac{1}{2}p^{2}(t)+V(x(t))-E)]}.\label{eq:KN}
\end{eqnarray}
let us first integrate over $p$ to get the Lagrangian form of the
path integral
\[
K(x_{1};x_{0})=\int\frac{[dN]}{\sqrt{N}}[dx]e^{\frac{i}{\hbar}\int_{0}^{1}dt[N(t)^{-1}\frac{1}{2}\dot{x}^{2}-N(t)(V-E)]}.
\]

From \eqref{eq:NRS} we have the following equations of motion:
\begin{eqnarray}
\delta_{N}S & = & 0\Rightarrow\frac{\dot{x}^{2}}{2}+N^{2}V=N^{2}E\label{eq:Neqn}\\
\delta_{x}S & = & 0\Rightarrow\ddot{x}=-N^{2}V'(x)\label{eq:xeqn}
\end{eqnarray}
Let us specialize to a linear potential (as in the mini-superspace
case with a positive CC) $V=V_{0}-\Lambda x,\,\Lambda>0$. The equation
for $x$ is solved by
\begin{equation}
x=\frac{1}{2}N^{2}\Lambda t^{2}+(x_{1}-x_{0}-\frac{1}{2}N^{2}\Lambda)t+x_{0}.\label{eq:xsoln}
\end{equation}
Evaluating \eqref{eq:Neqn} at $t=0$ and using \eqref{eq:xsoln}
gives a quartic equation for $N$ (after gauge fixing $\dot{N}=0$)
\[
\frac{\Lambda^{2}}{8}N^{4}-N^{2}\left(E-V_{0}+\frac{1}{2}\Lambda(x_{1}+x_{0})\right)+\frac{1}{2}(x_{1}-x_{0})^{2}=0
\]
which has the four solutions
\[
N=\pm\frac{\sqrt{2}}{\Lambda}[(E-V_{0}+\Lambda x_{1})^{1/2}\pm(E-V_{0}+\Lambda x_{0})^{1/2}]
\]
 corresponding to the four (in general complex) saddle points found
in \citep{Feldbrugge:2017kzv} for the mini-superspace case. In effect
these authors would have calculated \eqref{eq:KN} in the saddle point
approximation and decided which contour to integrate $N$ over and
hence which saddle points to pick up based on a Picard-Lefshetz analysis.
In fact it has been argued by these authors that one should integrate
only over positive $N$.

However the classical action is completely independent of these! Solving
\eqref{eq:Neqn} for $\dot{x}$ and substituting in \eqref{eq:NRS}
\begin{eqnarray*}
S & = & \int_{0}^{1}dt[N^{-1}N^{2}(E-V)+N(E-V)]=\int_{0}^{1}dt[2N(E-V)]\\
 & = & \int_{x_{0}}^{x_{1}}\frac{dx}{\dot{x}}2N(E-V)=\pm\int_{x_{0}}^{x_{1}}\frac{dx}{N\sqrt{2(E-V)}}N2(E-V)\\
 & = & \pm\int_{x_{0}}^{x_{1}}dx\sqrt{2(E-V)}
\end{eqnarray*}
This is independent of the saddle points for $N$ and gives for $K(x_{1};x_{0})$
as expected two different results as was the case with the wave function
calculation \eqref{eq:PsiWKB}.

In any case as we argued in the introduction the analog of the Feynman
kernel is not the physically interesting quantity to calculate. The
probability for tunneling is given by the squares of (or ratios of
squares of wave functions). Again the point is that we are looking
at energy eigenstates and asking what the probability of finding the
position of the particle (which of course is not a variable which
commutes with the Hamiltonian) on one or other side of the potential
barrier.

\subsection{Particle creation in an E-field.}

In \citep{Brown:1988kg} Brown and Teitelboim (BT) used a Euclidean
instanton to describe pair creation in an electric field (the Schwinger
process) in 1+1 dimensions, and then extended it to brane nucleation
in higher dimensions. We will just focus on the former in flat space
and ignore the dynamics of the E-M field. 

\begin{equation}
S=-m\int ds(-\eta_{\mu\nu}\dot{x}^{\mu}\dot{x}^{\nu})^{1/2}-e\int ds\dot{x}^{\mu}A_{\mu},\,\,\dot{x}^{\mu}\equiv\frac{dx^{\mu}}{ds}.\label{eq:partaction}
\end{equation}
Introduce a proper time metric factor $N(s)$ to write the action
in quadratic form and choose the gauge potential for a constant electric
field $E$ as $A_{\mu}=(Ex,0)$ so 
\begin{equation}
S=-\int ds\left\{ \frac{m}{2N(s)}(\dot{t}^{2}-\dot{x}^{2})+\frac{m}{2}N(s)\right\} -eE\int ds\dot{t}x.\label{eq:action2}
\end{equation}
In the canonical formalism we have conjugate momenta
\begin{eqnarray*}
\pi_{t} & = & -\frac{m}{N}\dot{t}-eEx,\,\,\pi_{x}=m\frac{\dot{x}}{N},\,\,\pi_{N}=0,
\end{eqnarray*}
and Hamiltonian

\begin{equation}
{\cal H}=N\left(-\frac{\pi_{t}^{2}}{2m}+\frac{\pi_{x}^{2}}{2m}-\frac{(eEx)^{2}}{2m}-\frac{e}{m}\pi_{t}Ex+\frac{m}{2}\right)\equiv NH.\label{eq:H}
\end{equation}
The dynamics is given by Hamilton's equations for the phase space
variables $\{\alpha\}=N,\pi_{N,}x,\pi_{x},t,\pi_{t}$,
\begin{equation}
\dot{\alpha}=\{\alpha,{\cal H}\},\label{eq:Heqnmotion}
\end{equation}
with the following relations/constraints.

Poisson Brackets:
\begin{equation}
\{N,\pi_{N}\}=\{x,\pi_{x}\}=\{t,\pi_{t}\}=1.\label{eq:PBs}
\end{equation}
Primary constraints: 
\begin{equation}
\pi_{N}\approx0.\label{eq:piN}
\end{equation}
Secondary constraints:
\begin{equation}
\dot{\pi}_{N}=\{\pi_{N},{\cal H}\}=-H\approx0.\label{eq:H0}
\end{equation}
The equations of motion are 
\begin{eqnarray}
\dot{N} & = & \{N,{\cal H}\}=0,\,\,\dot{t}=-\frac{N}{m}(\pi_{t}+eEx),\,\,\dot{x}=N\frac{\pi_{x}}{m},\nonumber \\
\dot{\pi}_{t} & = & 0,\,\,\dot{\pi}_{x}=N(eE)^{2}\frac{x}{m}.\label{eq:pitpixdot}
\end{eqnarray}
Since $\pi_{t}$ is constant let us choose 
\begin{equation}
\pi_{t}=\pi_{t0}=0.\label{eq:pit0}
\end{equation}
 Also in the passage to QM $\pi_{q}\rightarrow-i\hbar\frac{\delta}{\delta q}$
etc. acting on the Schroedinger wave function $\Psi$. In particular
$\pi_{N}\approx0$ implies 
\[
\frac{\delta}{\delta N}\Psi=0,
\]
 and $H\approx0$ implies the ``Wheeler-DeWitt'' equation
\[
H(q,-i\hbar\frac{\delta}{\delta q})\Psi(q)=0.
\]
(We are assuming there is no boundary in space and hence no boundary
Hamiltonian.)

In the WKB approximation (ignoring the pre-factor) 
\begin{equation}
\Psi\propto e^{\frac{i}{\hbar}S_{_{cl}}}\label{eq:Psi}
\end{equation}
where the classical action (evaluated on a solution) is 
\begin{eqnarray}
S_{cl} & = & \int ds[\pi_{t}\dot{t}+\pi_{x}\dot{x}+\pi_{N}\dot{N}-{\cal H})\nonumber \\
 & = & \int\pi_{x}dx.\label{eq:Scl}
\end{eqnarray}
In the last step we used \eqref{eq:piN}\eqref{eq:pit0}. Again as
in the previous example we see that imposing the classical constraints
on the motion leaves the classical action completely independent of
the lapse $N$. Also from \eqref{eq:H0}\eqref{eq:pit0} we have solving
for $\pi_{x}$
\[
\pi_{x}=\pm|eE|\sqrt{x^{2}-\left(\frac{m}{eE}\right)^{2}}.
\]
 Evaluating the integral between the two turning points one gets (defining
$\gamma=m/|eE|$
\begin{eqnarray}
S_{cl} & = & \pm|eE|\int_{-\gamma}^{+\gamma}dx\sqrt{x^{2}-\gamma^{2}}\nonumber \\
 & = & \pm|eE|\left[\frac{1}{2}x\sqrt{x^{2}-\gamma^{2}}-\frac{1}{2}\gamma^{2}\ln(x+\sqrt{x^{2}-\gamma^{2}}\right]_{-\gamma}^{+\gamma}\nonumber \\
 & = & \pm\frac{i\pi m^{2}}{2|eE|}(2n+1),\,\,n\in{\cal Z}^{\ge}.\label{eq:Scl-1}
\end{eqnarray}
This gives a probability (from the dominant solution $n=0$), 
\begin{equation}
P\propto|\Psi|^{2}\propto e^{-\frac{\pi m^{2}}{|eE|}}\label{eq:P}
\end{equation}
In agreement with Brown and Teitelboim's instanton calculation (and
the leading term in Schwinger's calculation). Note that here we have
rejected the possible positive sign in the exponent on physical grounds
since otherwise we would have $P$ rising with decreasing electric
field $E$! In other words there is no way that one can get the right
sign from the formalism.

Let us redo the calculation in the Lagrangian formulation with the
lapse gauge fixed to $N(s)=N$ (a constant as in \citep{Feldbrugge:2017kzv}).
The action is again given by \eqref{eq:action2} and the Lagrangian
equations of motion are,
\[
\ddot{t}=\alpha\dot{x},\,\,\ddot{x}=\alpha\dot{t},\,\,\alpha\equiv-\frac{eEN}{m},
\]
from varying w.r.t. $x^{\mu}$, and from varying w.r.t. $N$,
\[
-\frac{m}{2N^{2}}(\dot{t}^{2}-\dot{x}^{2})+\frac{m}{2}=0.
\]
Define $x^{\pm}=t\pm x$ so that the above equations become
\[
\ddot{x}^{\pm}=\pm\alpha\dot{x}^{\pm},\,\,N^{2}=\dot{x}^{+}\dot{x}^{-}.
\]
With appropriate choice of initial conditions we have the solutions
\begin{equation}
\dot{x}^{\pm}=\dot{x}_{0}^{\pm}e^{\pm\alpha s},\,\,x^{\pm}=\pm\frac{\dot{x}_{0}^{\pm}}{\alpha}e^{\pm\alpha s},\,\,N^{2}=\dot{x}_{0}^{+}\dot{x}_{0}^{-}.\label{eq:solns}
\end{equation}
The equation for the orbit is:
\begin{equation}
t^{2}-x^{2}=-\frac{\dot{x}_{0}^{+}\dot{x}_{0}^{-}}{\alpha^{2}}=-\frac{m^{2}}{(eE)^{2}},\label{eq:orbit}
\end{equation}
where in the last step we used the last equation of \eqref{eq:solns}.
Note that at $t=0$ we have 
\begin{equation}
x(t=0)=\pm\frac{m}{eE}\equiv\pm\gamma.\label{eq:xt0}
\end{equation}
 In the Brown-Teitelboim description of pair creation, the particle
on the left propagates backwards in time (anti-particle) $t$ while
the particle on the right propagates forward. This is to be interpreted
as pair creation at time $t=0$ at the points given by \eqref{eq:xt0}.
Along the space-like tunnneling trajectory $\dot{x}_{0}^{+}\dot{x}_{0}^{-}=-|\dot{x}_{0}^{+}\dot{x}_{0}^{-}|,$
which implies that the saddle points for the integration over $N$
are pure imaginary, $N=\pm i\sqrt{|\dot{x}_{0}^{+}\dot{x}_{0}^{-}|}.$
But the point is the $E$ dependence of the action at the classical
solutions comes only from the second term of \eqref{eq:action2} which
is independent of which solution for $N$ one chooses. Evaluating
the last term in the action over the tunneling trajectory gives an
imaginary part to the action from the term,
\[
eE\int_{x=-\gamma}^{x=\gamma}ds\dot{t}x=-\int_{-\gamma}^{\gamma}dxt(x)=\mp\int_{-\gamma}^{\gamma}dx\sqrt{x^{2}-\gamma^{2}},
\]
which is the same as before i.e. \eqref{eq:Scl-1} and hence gives
the same probability for pair creation \eqref{eq:P}. 

The point of this exercise was to show that the picking one or other
solution for $N$ does not resolve the sign ambiguity that we had
in the Hamiltonian discussion. As there this ambiguity comes from
the two solutions for $t$, which in that case was from solving the
Hamiltonian constraint while here it comes from solving the orbit
equation \eqref{eq:orbit}. 

Felbrugge et al \citep{Feldbrugge:2017fcc} claim to be able to resolve
this difference by computing the Feynman kernel for propagation from
$t_{0},x_{0}$ to $t_{1},x_{1}$ (as in the case of the simple QM
calculation of the previous example). However the reason for this
resolution is the imposition of a certain ``causality'' criterion
on the integral over $N$. However the forward direction of time is
ambiguous in this one-particle quantum mechanics discussion of what
is essentially a field theoretic process. Does one regard this as
an electron traveling first backward in time and then forward or a
positron traveling forward in time and then backward or vice-versa.
In the field theoretic argument on the other hand unitarity can resolve
this ambiguity as in the original Schwinger calculation.

\section{Mini-superspace }

\subsection{Background wave function}

Let us now discuss mini-superspace in the Hamiltonian formalism. The
action is given by (after setting $M_{P}^{-1}=8\pi G=1$) 
\begin{equation}
S=\int_{0}^{1}dt\left(-N^{-1}3a\dot{a}^{2}+(3ka-a^{3}\Lambda)N\right).\label{eq:MSaction}
\end{equation}
Note for future reference that the actual action (in 3+1 dimensions)
has a factor of the unit three sphere volume and is $2\pi^{2}S$.
Also $\dot{x}\equiv\frac{dx}{dt}$. Change variable to $q=a^{2}$
$N\rightarrow N/a$:
\[
S=\int dt\left(-\frac{3}{4N}\dot{q}^{2}+N(3k-\Lambda q)\right),
\]
\begin{equation}
\pi_{N}=0,\,\,\pi_{q}=-\frac{3}{2N}\dot{q},\,\,\dot{q}=-\frac{2N\pi_{q}}{3},\label{eq:pi}
\end{equation}
\begin{equation}
{\cal H}=N\left(-\frac{\pi_{q}^{2}}{3}+(\Lambda q-3k)\right)\equiv NH.\label{eq:H-1}
\end{equation}

The primary and secondary constraints are,
\[
\pi_{N}=0,\,\,\dot{\pi}_{N}=\{\pi_{N},{\cal H}\}=H\approx0.
\]
So on a classical trajectory
\[
\pi_{q}^{2}=3(\Lambda q-3k),\,\,\pi_{q}=\pm3\sqrt{\frac{\Lambda}{3}q-k}.
\]
TheWheeler-DeWitt equation for the system is obtained by putting $\pi_{q}\rightarrow-i\hbar\partial/\partial q$
in the Hamiltonian constraint and is 
\begin{equation}
\left\{ +\frac{\hbar^{2}}{3}\frac{\partial^{2}}{\partial q^{2}}+(\Lambda q-3)\right\} \Psi=0.\label{eq:WdW0}
\end{equation}
 Consider tunneling to a deSitter space with $\Lambda>0,\,\,k=1$.
The classical action is
\begin{eqnarray*}
\int_{q_{0}}^{q_{1}}\pi_{q}dq & = & \pm3\int_{q_{0}}^{q_{1}}dq\sqrt{\frac{\Lambda}{3}q-k}\\
 & = & \mp\frac{6i}{\Lambda}\left[\left(1-\frac{\Lambda}{3}q_{1}\right)^{3/2}-\left(1-\frac{\Lambda}{3}q_{0}\right)^{3/2}\right].
\end{eqnarray*}
This is pure imaginary for ``under the barrier'' propagation $q_{0},q_{1}<\frac{3}{\Lambda}$.
In the semi-classical approximation (and ignoring the fluctuations)
this gives the transition amplitude (ignoring pre-factors!)
\begin{eqnarray}
K(a_{1};a_{0}) & \sim & \exp\left[i2\pi^{2}\int_{q_{0}}^{q_{1}}\pi_{q}dq\right]\nonumber \\
 & = & \exp\left[\pm\frac{12\pi^{2}}{\Lambda}\left\{ \left(1-\frac{\Lambda}{3}a_{1}^{2}\right)^{3/2}-\left(1-\frac{\Lambda}{3}a_{0}^{2}\right)^{3/2}\right\} \right]\label{eq:K}
\end{eqnarray}
Thus again the Hamiltonian analysis shows that both signs are allowed,
i.e. both HH and T are valid solutions. The two signs come again from
the fact that the $H$ constraint is quadratic in $\pi$. As in the
case of the particle in an E-field the sign has to be chosen on physical
grounds. One might argue as in the latter case that the upper sign
(T) is physically more plausible. Actually the probability to find
a scale factor $a_{1}$ , 
\begin{equation}
P(a_{1})\sim|\Psi(a_{1})|^{2}\label{eq:P-1}
\end{equation}
 depends on the initial wave function since
\[
\Psi(a_{1})=\int da_{0}\mu(a_{0)}K(a_{1};a_{0})\Psi_{0}[a_{0}].
\]
 If we take the ``initial'' mini-superspace universe to be an eigenstate
of the scale factor with zero scale factor (``nothing''), $\Psi_{0}\propto\delta(a_{0})$.
So 
\begin{equation}
P(a_{1})\sim\exp\left[\mp\frac{24\pi^{2}}{\Lambda}\left\{ 1-\left(1-\frac{\Lambda}{3}a_{1}^{2}\right)^{3/2}\right\} \right].\label{eq:P-2}
\end{equation}
The upper sign gives a falling probability with increasing $a_{1}$
(tunneling (T) -Vilenkin) while the lower sign a rising probability
(HH).

We should note in passing that this wave function is an asymptotic
expression that is valid only in the region that is not only inside
but also far from the turning point. In particular it cannot be used
at the turning point $a=\sqrt{\frac{3}{\Lambda}}$ itself. We shall
discuss this further in the next subsection.

The authors of \citep{Feldbrugge:2017kzv} claim to fix the sign ambiguity
by first integrating over $q$ and then doing an integral over positive
$N$ and picking what they claim is the correct saddle point. Somehow
this seems to imply that the above calculation must fail for the ``wrong''
sign! As in the case of the non-relativistic particle and that of
the charged particle in an E field, it is clear from this analysis
that this ambiguity cannot be fixed by an extraneous notion of causality.

\subsection{Hartle-Hawking or Tunneling}

We argued in the previous subsections that the choice of the overall
sign of the exponent in \eqref{eq:K}\eqref{eq:P-2} is not determined
by any mathematical consistency argument, but maybe fixed by physical
considerations. To discuss this we need to match each under the barrier
(real) wave function to the appropriate (linear combination of) oscillating
wave functions. This is done by the standard WKB matching conditions.

In fact if one ignores the fluctuations, one simply has a linear potential
and the exact solution is well-known (eg. \citep{Merzbacher:1998}).
Thus defining 
\begin{equation}
z=\left(\frac{3}{\hbar^{2}\Lambda^{2}}\right)^{1/3}(3-\Lambda q),\label{eq:zdef}
\end{equation}
the Wheeler-DeWitt equation \eqref{eq:WdW0} becomes the Airy function
equation
\[
\frac{d^{2}f(z)}{dz^{2}}-zf(z)=0.
\]
The exact solution for the wave function is thus
\begin{equation}
\Psi(q)=A{\rm Ai}(z(q)+B{\rm Bi}(z(q).\label{eq:Psiexact}
\end{equation}
The asymptotic behavior of the Airy functions in the classical region
$z\ll-1$ i.e. $a\gg\sqrt{\frac{3}{\Lambda}}$ is
\begin{eqnarray}
{\rm Ai(z)} & \sim & \pi^{-1/2}|z|^{-1/4}\cos(\frac{2}{3}|z|^{3/2}-\frac{\pi}{4}),\label{eq:Aiz-}\\
{\rm Bi}(z) & \sim & -\pi^{-1/2}|z|^{-1/4}\sin(\frac{2}{3}|z|^{3/2}-\frac{\pi}{4}).\label{eq:Biz-}
\end{eqnarray}
On the other hand in the non-classical regime $z\gg1$ (which of course
can exist only for a very small cosmological constant $\Lambda\ll1$),
one has

\begin{eqnarray}
{\rm Ai(z)} & \sim & \frac{1}{2}\pi^{-1/2}|z|^{-1/4}\exp\left\{ -\frac{2}{3}z{}^{3/2}\right\} ,\label{eq:Aiz+}\\
{\rm Bi}(z) & \sim & \pi^{-1/2}|z|^{-1/4}\exp\left\{ \frac{2}{3}z{}^{3/2}\right\} .\label{eq:Biz+}
\end{eqnarray}
Suppose that we interpret the observed condition of an expanding universe
to mean that $\Psi$ should be an eigenstate of momentum
\[
-i\hbar\frac{d}{dq}\Psi=p\Psi,
\]
where by identifying the eigenvalue with the classical momentum \eqref{eq:pi}
we have
\[
p=\pi_{q}^{{\rm cl}}=-\frac{3}{2N}\dot{q}<0,
\]
for an expanding universe. Thus we need 
\begin{eqnarray}
\Psi_{{\rm out}} & \sim & e^{-i\frac{6}{\hbar\Lambda}\left(\frac{\Lambda q}{3}-1\right)^{3/2}+i\frac{\pi}{4}}\label{eq:expandingPsi}\\
 & \propto & {\rm Ai(z(q))+i{\rm Bi(z(q))}},\nonumber 
\end{eqnarray}
which gives $p=-3\left(\frac{\Lambda q}{3}-1\right)^{1/2}<0$ and
hence $\dot{a}>0$. Thus the corresponding under the barrier (i.e.
for $\frac{\Lambda q}{3}<1)$ wave function is 
\begin{eqnarray}
\Psi_{{\rm under}}(q) & = & A\pi^{-1/2}\left[\left(\frac{3}{\hbar^{2}\Lambda^{2}}\right)^{1/3}(3-\Lambda q)\right]^{-1/4}\times\nonumber \\
 &  & \left[\frac{1}{2}e^{-\frac{6}{\hbar\Lambda}\left(1-\frac{\Lambda q}{3}\right)^{3/2}}+ie^{\frac{6}{\hbar\Lambda}\left(1-\frac{\Lambda q}{3}\right)^{3/2}}\right].\label{eq:Psiunder}
\end{eqnarray}
This is Vilenkin's tunneling wave function proposal. The condition
that the observed universe is expanding (much as we observe electrons
coming out of the nucleus in $\beta$- decay), is used to impose this
outgoing boundary condition. Note that we have not normalized these
functions. Indeed it is not clear to us whether wave functionals in
a QFT (leave alone in quantum gravity) are even in principle normalizable.
However the relative probability may still make sense. So the relevant
probability for tunneling from ``nothing'' is 
\begin{equation}
P_{{\rm Tunnel}}(a;0)\equiv\frac{|\Psi_{{\rm out}}(q)|^{2}}{|\Psi_{{\rm under}}(0)|^{2}}\sim\exp\left(-\frac{12}{\hbar\Lambda}\right)\label{eq:Ptunnel}
\end{equation}

On the other hand Hartle and Hawking argued that the under the barrier
wave function is given by Euclidean quantum gravity. This amounts
to demanding that the first term in \eqref{eq:Psiunder} be the only
allowed wave function (i.e. we must pick the solution ${\rm Ai}$).
Then in the large $q$ regime 
\begin{equation}
\Psi\sim{\rm Ai}(z(q))\rightarrow\cos\left[\frac{6}{\hbar\Lambda}\left(\frac{\Lambda q}{3}-1\right)^{3/2}-\frac{\pi}{4}\right],\label{eq:HHclassical}
\end{equation}
corresponding to a superposition of an expanding and contracting universe.
In other words one does not have a classical expanding background
and needs to appeal to some sort of decoherence argument to account
for the observed expanding universe. In this case 
\begin{equation}
P_{{\rm HH}}(a;0)\equiv\frac{|\Psi_{{\rm out}}(q)|^{2}}{|\Psi_{{\rm under}}(0)|^{2}}\sim\exp\left(\frac{12}{\hbar\Lambda}\right)\label{eq:PHH}
\end{equation}

All this is well-known and is included here for completeness and to
set the stage for identifying the fluctuation spectrum around these
solutions.

\subsection{Fluctuations\label{sub:Fluctuations} }

Now one might ask whether the ambiguity as to which wave function
is physical might be fixed by the inclusion of tensor perturbations.
According to \citep{Feldbrugge:2017fcc,Feldbrugge:2017mbc}, the leading
order calculation of tensor perturbations has the wrong (inverse Gaussian)
sign for both wave functions, implying that there is no smooth beginning
to the universe in either case. This is contrary to the claim of \citep{DiazDorronsoro:2017hti}
(and citations therein) that HH leads to Gaussian fluctuations, while
the tunneling wave function is unstable to quadratic fluctuations.

The tensor modes are expanded in $S_{3}$ spherical harmonics labelled
by integers $l,m,n$. Suppressing the last two indices the action
for a mode $\phi_{l}$ in the deSitter background is to quadratic
order \citep{Feldbrugge:2017mbc} 
\begin{equation}
S_{l}=\frac{1}{2}\int dt\left[q^{2}\frac{\dot{\phi}_{l}^{2}}{N}-Nl(l+2)\phi_{l}^{2}\right].\label{eq:Sphi}
\end{equation}
So (dropping also the subscript $l$), 
\begin{equation}
\pi_{\phi}=q^{2}\frac{\dot{\phi}}{N},\,\,\dot{\phi}=\frac{N\pi_{\phi}}{q^{2}},\label{eq:piphi}
\end{equation}
giving the total Hamiltonian 
\begin{equation}
{\cal H}=NH=N\left\{ -\frac{1}{3}\pi_{q}^{2}+(\Lambda q-3)+\frac{\pi_{\phi}^{2}}{2q^{2}}+\frac{1}{2}l(l+2)\phi^{2}\right\} .\label{eq:Htotal}
\end{equation}
The WdW equation is: ($\pi_{q}\rightarrow-i\hbar\partial/\partial q,\,\,\pi_{\phi}\rightarrow-i\hbar\partial/\partial/\partial\phi$)
\begin{equation}
\left\{ +\frac{\hbar^{2}}{3}\frac{\partial^{2}}{\partial q^{2}}+(\Lambda q-3)-\frac{\hbar^{2}}{2q^{2}}\frac{\partial^{2}}{\partial\phi^{2}}+\frac{1}{2}l(l+2)\phi^{2}\right\} \Psi=0.\label{eq:WdW}
\end{equation}
Building on the solution in the absence of $\phi$, we try
\begin{equation}
\Psi[q,\phi]\sim\exp\left\{ \frac{1}{\hbar}\left(\frac{6}{\Lambda}c_{1}(1-\frac{\Lambda}{3}q)^{3/2}+c_{2}\frac{1}{2}q\phi^{2}\sigma(q)\right)\right\} ,\label{eq:Psi-1}
\end{equation}
where $c_{1}=\pm1,\,c_{2}=\pm1$ and $\sigma_{l}$ is a function that
is to be determined by \eqref{eq:WdW}. Computing derivatives we get;

\[
\partial_{q}\Psi=\frac{1}{\hbar}\left\{ -c_{1}3(1-\frac{\Lambda}{3}q)^{1/2}+c_{2}\frac{1}{2}\phi^{2}(\sigma(q)+q\sigma'(q))\right\} \Psi,
\]
\begin{equation}
\frac{\hbar^{2}}{3}\partial_{q}^{2}\Psi=\left\{ (3-\Lambda q)-c_{1}c_{2}\phi^{2}(\sigma(q)+q\sigma'(q))(1-\frac{\Lambda}{3}q)^{1/2}+O(\phi^{4})+O(\hbar)\right\} \Psi,\label{eq:d2Psi}
\end{equation}
\[
-\frac{\hbar^{2}}{2q^{2}}\partial_{\phi}^{2}\Psi=[-\frac{1}{2}\phi^{2}\sigma^{2}(q)+O(\hbar)]\Psi.
\]
Thus ignoring the $\phi^{4}$ and $\hbar$ corrections the WdW equation
\eqref{eq:WdW} becomes
\begin{equation}
[-c_{1}c_{2}\phi^{2}(\sigma(q)+q\sigma'(q))(1-\frac{\Lambda}{3}q)^{1/2}-\frac{1}{2}\phi^{2}\sigma^{2}(q)+\frac{1}{2}l(l+2)\phi^{2}]\Psi=0.\label{eq:WdW1}
\end{equation}
Thus $\sigma_{l}$ is required to be a solution of 
\begin{equation}
-c_{1}c_{2}(\sigma(q)+q\sigma'(q))(1-\frac{\Lambda}{3}q)^{1/2}-\frac{1}{2}\sigma^{2}(q)+\frac{1}{2}l(l+2)=0.\label{eq:sigmaeqn}
\end{equation}
Putting
\[
\sigma(q)=\frac{g_{l}}{(1-\frac{\Lambda}{3}q)^{1/2}+f_{l}},
\]
we see that \eqref{eq:sigmaeqn} is satisfied with 
\begin{equation}
f_{l}=\pm(l+1),\,\,g_{l}=c_{1}c_{2}l(l+2).\label{eq:flgl}
\end{equation}
Thus (since $c_{2}^{2}=1$) we have the four solutions for the wave
function (up to terms $O(\phi^{4}),$ $O(\hbar)$ in the exponent)
\begin{eqnarray}
\Psi & \sim & \exp\left\{ \frac{c_{1}}{\hbar}\left(\frac{6}{\Lambda}(1-\frac{\Lambda}{3}q)^{3/2}+\frac{1}{2}q\phi^{2}\frac{l(l+2)}{(1-\frac{\Lambda}{3}q)^{1/2}\pm(l+1)}\right)\right\} .\label{eq:Psi-2}\\
 & = & \exp\left\{ \frac{c_{1}}{\hbar}\left(\frac{6}{\Lambda}(1-\frac{\Lambda}{3}q)^{3/2}+\frac{1}{2}q\phi^{2}\frac{l(l+2)\left(-(1-\frac{\Lambda}{3}q)^{1/2}\pm(l+1)\right)}{\frac{\Lambda}{3}q+l(l+2)}\right)\right\} .
\end{eqnarray}
Similarly we have four possible solutions in the classically allowed
region,
\begin{eqnarray}
\Psi & \sim & \exp\left\{ \frac{c_{1}}{\hbar}\left(\frac{6}{\Lambda}i(\frac{\Lambda}{3}q-1)^{3/2}-\frac{1}{2}q\phi^{2}\frac{l(l+2)}{i(\frac{\Lambda}{3}q-1)^{1/2}\pm(l+1)}\right)\right\} \nonumber \\
 & = & \exp\left\{ \frac{c_{1}}{\hbar}\left(\frac{6}{\Lambda}i(\frac{\Lambda}{3}q-1)^{3/2}+\frac{1}{2}q\phi^{2}\frac{l(l+2)\left(i(\frac{\Lambda}{3}q-1)^{1/2}\mp(l+1)\right)}{\frac{\Lambda}{3}q+l(l+2)}\right)\right\} .\label{eq:Psi-2out}
\end{eqnarray}

Note that these four solutions are in agreement with the solutions
for the classical action including quadratic fluctuations given in
eqn. (30) of \citep{Feldbrugge:2017mbc}. It is clear that two of
the four solutions (both under and outside the barrier) give Gaussian
fluctuations while two give inverse Gaussian fluctuations. Obviously
the general solution which would be a linear superposition of these
four solutions will necessarily have components which would invalidate
the smooth background which is the starting assumption of this analysis.
Feldebrugge et al claim that their path integral argument (essentially
involving an ordinary integral over the lapse $N$) necessarily includes
a non-Gaussian component in both the T and HH solutions. In other
words depending on the choice of contour (i.e. with a given sign for
$c_{1}$) one needs to include both $\pm$ signs (corresponding to
the two signs for $f_{l}$ in eqn. \eqref{eq:flgl}) in the solution. 

Nevertheless as we will discuss below once one uses the matching conditions
to evaluate the wave function in the classical regime, it will be
seen that we get different linear combinations of the under barrier
wave functions. In fact one can choose to pick those linear combinations
which admit only Gaussian fluctuations (with the appropriate choice
of sign in $f_{l}$ in equation \eqref{eq:flgl}). This can be done
for both the tunneling wave function as well as the Hartle-Hawking
wave function. 

Before we discuss this we would like to point out that the wave function
necessarily contains non-Gaussian fluctuations. Even ignoring $\hbar$
corrections there are $O(\phi^{4})$ terms 
\[
\frac{1}{12}(\sigma(q)+q\sigma'(q))^{2}\phi^{4}
\]
 in \eqref{eq:d2Psi} which imply that the quadratic (in $\phi$)
term in the log of the wave function $\Psi$ in \eqref{eq:Psi-1}
needs to include a quartic term in $\phi$! This does not necessarily
imply measurable non-Gaussianities in the CMB spectrum since it is
possible that they are significant only at points in field space where
the perturbative analysis of this paper, namely the expansion in fluctuations
$\phi$, breaks down \footnote{I thank an anonymous referee for emphasizing this.}.

\subsection{Wave function with fluctuations}

Using the matching conditions for the zeroth order (in fluctuations)
wave functions given by the asymptotic behavior of the Airy functions
\eqref{eq:Aiz-}\eqref{eq:Biz-}\eqref{eq:Aiz+}\eqref{eq:Biz+},
and the correlation that we found between the sign of the fluctuations
and the particular solution to the zeroth order equation \eqref{eq:Psi-2}
(see also \eqref{eq:sigma1<}\eqref{eq:sigma1>}), we can now write
down the wave function in the classically forbidden and allowed regions
for the asymptotic regimes and for small $\phi$ fluctuations. To
simplify the formulae let us first make the following definitions:
\[
\lambda(q)\equiv\frac{6}{\Lambda}\vert\left(1-\frac{\Lambda q}{3}\right)\vert^{3/2},\,\,\Delta\equiv\frac{\Lambda}{3}q+l(l+2)>0
\]
For $\frac{\Lambda q}{3}\ll1$ we have
\begin{eqnarray}
\Psi_{{\rm in}}(q,\phi) & = & A_{+}e^{-\frac{1}{\hbar}\left\{ \lambda(q)+\frac{1}{2}q\phi^{2}\frac{l(l+2)\left(-(1-\frac{\Lambda}{3}q)^{1/2}+(l+1)\right)}{\Delta}\right\} }+A_{-}e^{-\frac{1}{\hbar}\left\{ \lambda(q)+\frac{1}{2}q\phi^{2}\frac{l(l+2)\left(-(1-\frac{\Lambda}{3}q)^{1/2}-(l+1)\right)}{\Delta}\right\} }\nonumber \\
 &  & +B_{+}e^{+\frac{1}{\hbar}\left\{ \lambda(q)+\frac{1}{2}q\phi^{2}\frac{l(l+2)\left(-(1-\frac{\Lambda}{3}q)^{1/2}+(l+1)\right)}{\Delta}\right\} }+B_{-}e^{+\frac{1}{\hbar}\left\{ \lambda(q)+\frac{1}{2}q\phi^{2}\frac{l(l+2)\left(-(1-\frac{\Lambda}{3}q)^{1/2}-(l+1)\right)}{\Delta}\right\} }\label{eq:psi-z-}
\end{eqnarray}
Note that the $A_{+}$ and $B_{-}$ terms have Gaussian fluctuations
while the $A_{-}$ and $B_{+}$ terms have inverse Gaussian fluctuations.
Hence one might choose the solution with 
\begin{equation}
A_{-}=B_{+}=0,\label{eq:incondition}
\end{equation}
 in order to avoid quadratic instabilities. Of course this is not
a guarantee that the solutions are indeed stable since we have nothing
to say about the sign of higher order fluctuations! 

In the asymptotic classical regime $\Lambda q/3\gg1$ the general
solution may be written as (we include a constant phase factor for
later convenience) 
\begin{eqnarray}
\Psi_{{\rm out}}(q,\phi) & = & C_{+}^{-}e^{\frac{1}{\hbar}\left\{ i\lambda(q)-\frac{q}{2}\phi^{2}\frac{l(l+2)}{D_{+}}\right\} -i\frac{\pi}{4}}+C_{-}^{+}e^{-\frac{1}{\hbar}\left\{ i\lambda(q)-\frac{q}{2}\phi^{2}\frac{l(l+2)}{D_{+}}\right\} +i\frac{\pi}{4}}\nonumber \\
 &  & +C_{+}^{+}e^{\frac{1}{\hbar}\left\{ i\lambda(q)-\frac{q}{2}\phi^{2}\frac{l(l+2)}{D_{-}}\right\} -i\frac{\pi}{4}}+C_{-}^{-}e^{-\frac{1}{\hbar}\left\{ i\lambda(q)-\frac{q}{2}\phi^{2}\frac{l(l+2)}{D_{-}}\right\} +i\frac{\pi}{4}}\label{eq:psi-z+}
\end{eqnarray}
with 
\[
\frac{1}{D_{\pm}}=\frac{-i(\Lambda q/3-1)^{1/2}\pm(l+1)}{\Delta}.
\]
In the above eqn.\eqref{eq:psi-z+} as well as in \eqref{eq:psi-z-}
the particular combination of the fluctuation term $\propto\phi^{2}$
and the background term $\propto\lambda(q)$ is determined by the
Wheeler DeWitt equation as indicated in \eqref{eq:Psi-2out}. To avoid
having solutions which are unstable to quadratic fluctuations in the
classical region we need to set 
\begin{equation}
C_{-}^{+}=0,\,\,C_{+}^{+}=0.\label{eq:outcondition}
\end{equation}
Let us now impose the requirement of suppressed quadratic fluctuations
in both the inside and the outside wave functions, and then match
the two in the limit where $\phi=0$. In this case we may use the
matching conditions of the unperturbed theory \eqref{eq:Aiz-}-\eqref{eq:Biz+}
which imply 
\begin{eqnarray*}
\frac{1}{2}e^{-\frac{\lambda(q)}{\hbar}} & \longleftrightarrow & \frac{1}{2}\left\{ e^{\frac{i}{\hbar}\lambda(q)-i\frac{\pi}{4}}+e^{-\frac{i}{\hbar}\lambda(q)+i\frac{\pi}{4}}\right\} ,\\
e^{\frac{\lambda(q)}{\hbar}} & \longleftrightarrow & \frac{1}{2i}\left\{ e^{\frac{i}{\hbar}\lambda(q)-i\frac{\pi}{4}}-e^{-\frac{i}{\hbar}\lambda(q)+i\frac{\pi}{4}}\right\} ,
\end{eqnarray*}
where the LHS of each relation corresponds to a classically forbidden
region and the RHS to a classically allowed region solution. These
conditions then determine $C_{+}^{-}=A_{+}-\frac{1}{2i}B_{-},\,\,C_{-}^{-}=A_{+}+\frac{1}{2i}B_{-}.$
For the sake of clarity let us then write out explicitly the solutions
with suppressed Gaussians in the two regions\footnote{A similar analysis for the case of tunneling boundary conditions was
done more than 30 years ago by Vachaspati and Vilenkin\citep{Vachaspati:1988as}. }:
\begin{eqnarray}
\Psi_{{\rm in}}(q,\phi) & = & A_{+}e^{-\frac{1}{\hbar}\left\{ \lambda(q)+\frac{1}{2}q\phi^{2}\frac{l(l+2)\left(-(1-\frac{\Lambda}{3}q)^{1/2}+(l+1)\right)}{\Delta}\right\} }+B_{-}e^{+\frac{1}{\hbar}\left\{ \lambda(q)+\frac{1}{2}q\phi^{2}\frac{l(l+2)\left(-(1-\frac{\Lambda}{3}q)^{1/2}-(l+1)\right)}{\Delta}\right\} }\label{eq:PsiinGauss}\\
\Psi_{{\rm out}}(q,\phi) & = & \left(A_{+}-\frac{1}{2i}B_{-}\right)e^{\frac{1}{\hbar}\left\{ i\lambda(q)-\frac{q}{2}\phi^{2}\frac{l(l+2)}{D_{+}}\right\} -i\frac{\pi}{4}}+\left(A_{+}+\frac{1}{2i}B_{-}\right)e^{-\frac{1}{\hbar}\left\{ i\lambda(q)-\frac{q}{2}\phi^{2}\frac{l(l+2)}{D_{-}}\right\} +i\frac{\pi}{4}}\label{eq:PsioutGauss}
\end{eqnarray}

Consider now Vilenkin's tunneling wave function case. The boundary
condition here is that there is only an outgoing component in the
classical region corresponding to an expanding universe, which means
setting $A_{+}+B_{-}/2i=0$. Thus we get
\begin{equation}
\Psi_{{\rm out}}^{(T)}(q,\phi)=2A_{+}e^{-\frac{1}{\hbar}\left\{ i\lambda(q)-\frac{q}{2}\phi^{2}\frac{l(l+2)}{D_{-}}\right\} +i\frac{\pi}{4}}.\label{eq:PsiTout}
\end{equation}
This should be compared with $\Psi_{{\rm in}}$ ,
\[
\Psi_{{\rm in}}^{(T)}(q\rightarrow0)=A_{+}\left[e^{-\frac{6}{\hbar\Lambda}}-2ie^{\frac{6}{\hbar\Lambda}}\right].
\]
The probability of the universe emerging in a Bunch-Davies vacuum
relative to remaining in a state of nothing (i.e with zero sale factor
$a=\sqrt{q}=0$) is (after restoring the $2\pi^{2}$ factor which
we had dropped), 
\begin{equation}
P^{(T)}(q,\phi)=\frac{|\Psi_{{\rm out}}^{(T)}(q,\phi)|^{2}}{|\Psi_{{\rm in}}^{(T)}(q\rightarrow0)|^{2}}=e^{-24\pi^{2}/\hbar\Lambda}e^{-\frac{2\pi^{2}}{\hbar}q\phi^{2}\frac{l(l+1)(l+2)}{\Delta}}.\label{eq:PT}
\end{equation}
Now let us consider the Hartle-Hawking case. Here if we insist that
the boundary conditions are given by Euclidean quantum gravity we
would need the wave function to be real. In this case using \eqref{eq:outcondition}
in \eqref{eq:psi-z+} and imposing reality we get (we choose $\Im A_{+}=0$
so $\Re B_{-}=0$) 
\begin{equation}
\Psi_{{\rm out}}^{(HH)}=4A_{+}\cos\left[\frac{1}{\hbar}\left(\lambda(q)+\frac{q\phi^{2}}{2}(\Lambda q/3-1)^{1/2}\frac{l(l+2)}{\Delta}\right)-\frac{\pi}{4}\right]e^{-\frac{2\pi^{2}}{\hbar}\phi^{2}\frac{l(l+1)(l+2)}{2\Delta}},\label{eq:psiHHout}
\end{equation}
in agreement with eqn. 3.23 of \citep{DiazDorronsoro:2017hti}. Furthermore
the under the barrier solution in this case in the limit of zero scale
factor is,
\[
\Psi_{in}^{(HH)}(q\rightarrow0)=2A_{+}e^{-\frac{6}{\hbar\Lambda}}.
\]
 Thus we have (after restoring the factor of $2\pi^{2}$) 
\begin{eqnarray}
P^{(HH)}(q,\phi) & = & \frac{|\Psi_{{\rm out}}^{(HH)}(q,\phi)|^{2}}{|\Psi_{{\rm in}}^{(HH)}(q\rightarrow0)|^{2}}=e^{24\pi^{2}/\hbar\Lambda}e^{-\frac{2\pi^{2}}{\hbar}q\phi^{2}\frac{l(l+1)(l+2)}{\Delta}}\times\nonumber \\
 &  & 4\cos^{2}\left[\frac{2\pi^{2}}{\hbar}\left(\lambda(q)+\frac{q\phi^{2}}{2}(\Lambda q/3-1)^{1/2}\frac{l(l+2)}{\Delta}\right)-\frac{\pi}{4}\right]\label{eq:PHH-1}
\end{eqnarray}
The Hartle-Hawking wave function is time symmetric between an expanding
and a contracting universe. The observed universe is of course expanding
so somehow the two branches must decohere - in which case the the
only difference between the two probabilities is in the pre-factor
- with the tunneling case (as is well-known) favoring a larger cosmological
constant and the Hartle-Hawking case favoring a smaller CC.

\subsection{Solving the Hamilton-Jacobi equation beyond quadratic order}

In the previous subsection we just considered the quadratic fluctuations
around the mini-superspace HH and tunneling solutions. However as
we mentioned at the end of subsection \eqref{sub:Fluctuations} it
is clear that the wave function necessarily contains non-Gaussianities.

To investigate the solutions systematically it is convenient to write
\begin{equation}
\Psi[q,\phi]=e^{\frac{i}{\hbar}S[q,\phi]}\label{eq:Psi-3}
\end{equation}
Substituting in the WdW equation \eqref{eq:WdW} we have the Hamilton-Jacobi
(HJ) equation plus its quantum correction:
\begin{equation}
-\frac{1}{3}\left(\frac{\partial S}{\partial q}\right)^{2}+(\Lambda q-3)+\frac{1}{2q^{2}}\left(\frac{\partial S}{\partial\phi}\right)^{2}+\frac{1}{2}l(l+2)\phi^{2}+i\hbar\left(\frac{1}{3}\frac{\partial^{2}S}{\partial q^{2}}-\frac{1}{2q^{2}}\frac{\partial^{2}S}{\partial\phi^{2}}\right)=0\label{eq:HJ+}
\end{equation}
The limit $\hbar\rightarrow0$ give the classical Hamilton-Jacobi
equation. Ignoring the quantum correction we try a solution of the
form (as before an overall factor of $2\pi^{2}$ is understood)
\begin{equation}
iS[q,\phi]=c_{1}\left[\frac{6}{\Lambda}(1-\frac{\Lambda}{3}q)^{3/2}+q(\sum_{n=1}^{\infty}\frac{1}{2n!}\sigma_{n}(q)\phi^{2n})\right]\label{eq:Sexpn}
\end{equation}
The HJ equation then gives
\begin{eqnarray*}
-2(1-\frac{\Lambda}{3}q)^{1/2}\sum_{n=1}^{\infty}\frac{1}{2n!}(\sigma_{n}(q)+q\sigma'_{n}(q))\phi^{2n} & +\\
\left(\sum_{n=1}^{\infty}\frac{1}{2n!}(\sigma_{n}(q)+q\sigma'_{n}(q))\phi^{2n}\right)^{2} & - & \frac{1}{2}\left(\sum_{n=1}^{\infty}\frac{1}{(2n-1)!}\sigma_{n}(q)\phi^{2n-1}\right)^{2}\\
+\frac{1}{2}l(l+2)\phi^{2} & = & 0
\end{eqnarray*}
Equating powers of $\phi^{2}$ gives a set of recursion relations
which in principle can be solved iteratively to determine $\sigma_{n}$.
For instance from the coefficient of $\phi^{2}$ we got (see eqns
\eqref{eq:sigmaeqn} to \eqref{eq:flgl}), 
\[
-(1-\frac{\Lambda}{3}q)^{1/2}(\sigma_{1}(q)+q\sigma'_{1}(q))-\frac{1}{2}\sigma_{1}^{2}+\frac{1}{2}l(l+2)=0,
\]
which is solved by 
\begin{equation}
\sigma_{1}=\frac{l(l+2)}{(1-\frac{\Lambda}{3}q)^{1/2}\pm(l+1)}.\label{eq:sigma1<}
\end{equation}
From the coefficient of $\phi^{4}$ we also get 
\begin{equation}
\frac{2}{4!}(1-\frac{\Lambda}{3}q)^{1/2}(\sigma_{2}+q\sigma'_{2})+\frac{1}{6}\sigma_{1}\sigma_{2}=\frac{1}{4}(\sigma_{1}+q\sigma'_{1})^{2}.\label{eq:sigma2}
\end{equation}
Since $\sigma_{1}+q\sigma'_{1}\ne0$ this shows that $\sigma_{2}\ne0$.
Clearly the higher order terms will all be non-zero. Note also that
in the classical regime $\left(\frac{\Lambda}{3}q>1,\,\,a>a_{dS}\equiv\sqrt{\frac{3}{\Lambda}}\right)$
the sigma's are complex since as we remarked before:
\begin{equation}
\sigma_{1}=\frac{l(l+2)[\pm(l+1)-i\sqrt{\frac{\Lambda}{3}q-1}]}{\frac{\Lambda}{3}q-1+(l+1)^{2}},\,\,{\rm etc.}\label{eq:sigma1>}
\end{equation}
The above calculation seems to indicate that solution of the WdW equation
in the mini-superspace approximation has non-Gaussian terms. It is
however unclear whether all of them are suppressed under the same
conditions that enabled us to get suppressed Gaussian fluctuations.

In view of the importance of this question let us investigate this
further using a slightly different parametrization - i.e. the one
used by Vachaspati and Vilenkin \citep{Vachaspati:1988as}. First
we rewrite the WdW equation \eqref{eq:WdW} in terms of the scale
factor $a$ to get 
\begin{equation}
\left\{ +\hbar^{2}\left(\frac{\partial^{2}}{\partial a^{2}}+p\frac{\partial}{\partial a}\right)+12a^{4}(\Lambda a^{2}-3)-6\hbar^{2}\frac{\partial^{2}}{\partial\phi^{2}}+6l(l+2)a^{4}\phi^{2}\right\} \Psi=0\label{eq:WdW2}
\end{equation}
This is essentially (apart from a trivial normalization difference)
the equation analyzed in \citep{Vachaspati:1988as} (but for a given
mode $\phi\equiv\phi_{l}$) except that the direct substitution $q=a^{2}$
would have given us $p=-1$ in the above whereas they chose $p=+1$.
But this is just a difference in factor ordering which is only relevant
for the $O(\hbar)$ correction which we ignore. So let us work with
the latter choice for $p$. In this case we may substitute $x=\ln a$
to get 
\begin{eqnarray}
\left\{ +\hbar^{2}\frac{\partial^{2}}{\partial x^{2}}+-6\hbar^{2}\frac{\partial^{2}}{\partial\phi^{2}}-V(x,\phi)\right\} \Psi & = & 0\label{eq:WdWV}\\
V(x,\phi)=V_{0}(x)+V_{2}(x)\phi^{2} & \equiv & -12e^{4x}(\Lambda e^{2x}-3)-6l(l+2)e^{4x}\phi^{2}\label{eq:Pot}
\end{eqnarray}
We write the wave function as (we will only write out explicitly the
under the barrier growing mode) 
\begin{equation}
\Psi(x,\phi)=\exp\left[-S_{0}(x)-\frac{1}{2}S_{2}(x)\phi^{2}-\frac{1}{4}S_{4}(x)\phi^{4}+O(\phi^{6})\right].\label{eq:Psiphi4}
\end{equation}
Substituting in \eqref{eq:WdWV} we get a set of recursion relations
as before, namely (denoting by a prime derivation w.r.t. $x$)
\begin{eqnarray}
S_{0}'^{2} & = & V_{0}\label{eq:S0}\\
S_{0}'S_{2}'-S_{2}^{2} & = & V_{2}\label{eq:S2}\\
\frac{1}{2}S'_{0}S'_{4}+\frac{1}{4}\left(S'_{2}\right)^{2}-2S_{2}S_{4} & = & 0,\label{eq:S4}\\
\frac{1}{t}S_{0}'S_{2t}'+\sum_{r=1}^{t-1}\frac{S_{2r}'}{2r}\frac{S'_{2(t-r)}}{2(t-r)}-6\sum_{r=1}^{t}S_{2r}S_{2(t+1-r)} & = & 0\,\,{\rm for}\,t\ge2\label{eq:S2t}
\end{eqnarray}
Clearly all the coefficients $S_{2t}$ may be recursively determined
in principle. Solving the equation for $S_{4}$ for instance we get
\begin{equation}
S_{4}(x)=-\frac{1}{2}e^{24\int_{-\infty}^{x}dx'\frac{S_{2}}{S'_{0}}(x')}\int_{-\infty}^{x}dx'e^{-4\int_{-\infty}^{x'}dx''\frac{S_{2}}{S'_{0}}(x'')}\frac{\left(S'_{2}(x')\right)^{2}}{S'_{0}(x')}.\label{eq:S4soln}
\end{equation}
Once a choice of $S_{0}'=\pm\sqrt{V_{0}}$ is made this equation (together
with the choice of solution for \eqref{eq:S2} which is essentially
given by $q\sigma_{1}$ - see \eqref{eq:sigma1<}\eqref{eq:sigma1>})
determines uniquely the coefficient $S_{4}$. In fact in the classically
forbidden region where $V_{0}>0$ and $S_{2}'$ is real, we see that
for the choice $S_{0}'=-\sqrt{V_{0}}$ (giving $S_{0}=+2\sqrt{3}(3-\Lambda a^{2})^{3/2}$),
we have $S_{4}>0$ corresponding to suppressed quartic fluctuations
in the rising under the barrier wave function, but the choice $S'_{0}=+\sqrt{V_{0}}$
gives growing quartic fluctuations signaling an instability in the
falling under the barrier wave function. 

What does this imply for the tunneling vs the Hartle-Hawking wave
functions? For the latter case (since we can set $B_{-}=0$) we have
suppressed quartic fluctuations. However for the former case since
$B_{-}$ is necessarily non-zero, we would have unsuppressed quartic
fluctuations in the under the barrier wave function. If this sign
persists for all higher order (i.e. $\phi^{2t}$) terms then clearly
the requirement of suppressed fluctuations will favor the HH wave
function. Unfortunately (as can be seen from \eqref{eq:S2t}) analyzing
the higher order (i.e. $2t\ge6)$ fluctuations is not straightforward,
so at this point we cannot say anything about these, and a definite
conclusion regarding these alternative solutions to the problem of
tunneling from 'nothing' cannot be made. 

While we cannot say anything about the sign of the higher order (i.e..
beyond quartic order) fluctuations the recursion relations \eqref{eq:sigma2}\eqref{eq:S4}\eqref{eq:S4soln}
(barring some unlikely cancellations) imply that this framework may
in general lead to non-Gaussianities. However as we observed earlier
this does not necessarily imply a breakdown of the Bunch-Davies vacuum
ansatz since it may be the case that these quartic (and higher order)
corrections become significant only at values of $\phi$ where our
expansion breaks down.

\section{Conclusions}

We have argued that there is no particular advantage to the saddle
point (Picard-Lefshetz) method in solving the WdW equation in the
minisuperspace truncation. Contrary to the work of Feldebrugge et
al \citep{Feldbrugge:2017fcc,Feldbrugge:2017kzv,Feldbrugge:2017mbc,Feldbrugge:2018gin}
which ascribes certain a priori criteria on properties of the integral
over the lapse and furthermore essentially compute a propagator rather
than a wave function, our point of view is essentially similar to
that of Diaz et al. \citep{DiazDorronsoro:2017hti,DiazDorronsoro:2018wro}
where the functional integral is treated as a means towards computing
solutions to the WdW equation. 

We've also argued that whether or not one picks the Hartle-Hawking
or the tunneling wave function is a matter of choosing one or the
other boundary condition, and so must be determined by some physical
input, and is not a matter of consistency of the saddle point method.
Furthermore we've shown how to include fluctuations around the mini-superspace
model and indeed the general solution does contain terms with inverse
Gaussian fluctuations as shown by Feldbrugge et al \citep{Feldbrugge:2017fcc,Feldbrugge:2017mbc}.
However we have demonstrated that one can choose to set some of the
arbitrary constants multiplying such terms to zero, so properly interpreted,
the quadratic fluctuations around both competing wave functions are
indeed of Gaussian form and are suppressed. 

On the other hand we have been unable to say anything definitive about
the sign of higher order fluctuations (apart from the quartic one
- which appears to have the wrong sign for the tunneling wave function)
hence it is difficult to rule out either formulation of quantum cosmology
based on purely theoretical grounds (as was the claim of \citep{Feldbrugge:2017fcc,Feldbrugge:2017mbc}.
However as argued at the end of the previous section - even if the
stability issue was settled it is hard to see how non-Gaussianities
could be suppressed (as required observationally) in this framework.

\section{Acknowledgements}

I wish to thank Cliff Burgess and Fernando Quevedo for discussions
and comments on the manuscript and Alex Vilenkin for a question on
a previous version of this paper which led me discuss the general
solution in detail. I also wish to acknowledge the hospitality of
the Abdus Salam ICTP where part of this work was done, and the receipt
of partial support from the Dean of the College of Arts and Sciences
at the University of Colorado, Boulder.

\bibliographystyle{apsrev}
\bibliography{myrefs}

\end{document}